\documentclass[12pt,a4paper]{article}
\usepackage{jcappub}
\usepackage{amsmath,amssymb,scalefnt}
\def\al{\alpha} 
\def\be{\beta}
\def\ga{\gamma}
\def\de{\delta}

\def\th{\theta}

\def\ka{\kappa}
\def\la{\lambda}

\def\si{\sigma}

\def\om{\omega}
\def\De{\Delta}
\def\Ga{\Gamma}

\def\La{\Lambda}

\def\Om{\Omega}

\newcommand{\ben}{\begin{equation}}
\newcommand{\een}{\end{equation}}
\newcommand{\bea}{\begin{eqnarray}}
\newcommand{\eea}{\end{eqnarray}}
\newcommand{\ba}{\begin{array}}
\newcommand{\ea}{\end{array}}
\newcommand{\bit}{\begin{itemize}}
\newcommand{\eit}{\end{itemize}}

\def\math{\mathsurround 0pt}
\def\oversim#1#2{\lower.5pt\vbox{\baselineskip0pt \lineskip-.5pt
 \ialign{$\math#1\hfil##\hfil$\crcr#2\crcr{\scriptstyle\sim}\crcr}}}

\def\Tr{\mathop{\rm Tr}\nolimits}

\newcommand{\vev}[1]{\left\langle#1\right\rangle}

\newcommand{\powspec}{\mathcal{P}}

\newcommand{\mpl}{m_\mathrm{P}}

\newcommand{\eqn}[1]{Eq.~(\ref{#1})}
\newcommand{\eqns}[2]{Eqs.~(\ref{#1}),~(\ref{#2})}

\newcommand{\reference}[1]{Ref.~\cite{#1}}

\newcommand{\abbrev}{\scalefont{.9}}
\newcommand{\susy}{{\abbrev SUSY}}
\newcommand{\sm}{{\abbrev SM}}
\newcommand{\rg}{{\abbrev RG}}

\newcommand{\lsp}{{\abbrev LSP}}
\newcommand{\nlsp}{{\abbrev NLSP}}
\newcommand{\mssm}{{\abbrev MSSM}}

\newcommand{\mhissm}{{\abbrev MHISSM}}

\newcommand{\cmssm}{{\abbrev CMSSM}}
\newcommand{\amsb}{{\abbrev AMSB}}
\newcommand{\gmsb}{{\abbrev GMSB}}

\newcommand{\samsb}{{\abbrev sAMSB}}

\newcommand{\bbn}{{\abbrev BBN}}
\newcommand{\cmb}{{\abbrev CMB}}

\def\ee{\end{equation}}

\def\frak#1#2{{\textstyle{\frac{#1}{#2}}}}
\def\half{\textstyle{\frac{1}{2}}}

\def\vev#1{\mathopen\langle #1\mathclose\rangle }
\def\nn{\nonumber\\}

\def\ga{\gamma} 
\def\de{\delta} 

\def \la{\lambda}
\def \La{\Lambda}
\def \th{\theta}

\def\sy{supersymmetry}

\def\phib{\overline{\phi}}
\def\Phib{\overline\Phi}

\def\mgrav{m_{\frac32}}
\def\msb{m_\textrm{sb}}
\def\barm{\bar{m}}
\def\tilh{\tilde{h}}

\title{\LARGE Consistent cosmology with Higgs thermal inflation in a minimal extension of the MSSM}

\author[a,c]{Mark~Hindmarsh}
\author[b] {D.~R.~Timothy~Jones}
\affiliation[a]{Dept.\ of Physics and Astronomy, University of Sussex, 
Brighton BN1 9QH, U.K.}
\affiliation[b]{Dept. of Mathematical Sciences,
University of Liverpool, Liverpool L69 3BX, U.K.}
\affiliation[c]{Helsinki Institute of Physics, P.O.\ Box 64, 00014 Helsinki University, Finland} 
\emailAdd{m.b.hindmarsh@sussex.ac.uk}
\emailAdd{drtj@liv.ac.uk}

\abstract{
We consider a class of supersymmetric inflation models, in which 
minimal gauged F-term hybrid inflation  is coupled renormalisably to
the minimal supersymmetric standard model (\mssm), with no extra
ingredients; we call this class the ``minimal hybrid inflationary
supersymmetric standard model'' (\mhissm).   The singlet inflaton couples
to the Higgs as well as the waterfall fields,  supplying the Higgs
$\mu$-term. We show how such models can exit inflation  to a vacuum
characterised by large Higgs vevs, whose vacuum  energy is controlled by
supersymmetry-breaking. The true ground state is reached after an
intervening  period of thermal inflation along the Higgs flat direction,
which has important consequences for  the cosmology of the F-term
inflation scenario. The scalar spectral index is reduced, with a value
of approximately 0.976 in the case where the inflaton  potential is
dominated by the 1-loop radiative corrections.  The reheat temperature
following thermal inflation is about $10^9$ GeV, which solves the
gravitino overclosure problem.  A Higgs condensate reduces the cosmic
string mass per unit length, rendering it compatible with the Cosmic
Microwave Background constraints without tuning the inflaton coupling. 
With the minimal U(1)$'$ gauge symmetry in the inflation sector, where 
one of the waterfall fields generates a
right-handed neutrino mass, we investigate the Higgs thermal inflation
scenario in three popular supersymmetry-breaking schemes: \amsb,
\gmsb\ and the \cmssm,  focusing on the implications for the gravitino
bound.   In \amsb\ enough gravitinos can be produced to account for the
observed dark matter abundance through  decays into neutralinos.
In \gmsb\ we find an upper bound on the  gravitino mass of about a TeV,  
while in the \cmssm\ the thermally generated gravitinos are
sub-dominant. When Big Bang Nucleosynthesis constraints are taken
into account,  the unstable gravitinos of \amsb\ and the \cmssm\ must
have a mass  O(10) TeV or greater, while in \gmsb\ we find an upper
bound on the gravitino mass of O(1) TeV. }

\begin{document}

\begin{flushright}
\vspace{-36pt}
{
HIP-2012-27/TH\\
LTH 958}
\end{flushright}

\maketitle

\section{Introduction}

Inflation is the accepted paradigm for the very early universe, thanks
to its power to account accurately for cosmological data in one simple
framework. However, it raises a number of theoretical problems,
principally the identity of the inflaton, the flatness of its potential,
and how it is coupled to the Standard Model.

A technically natural way of achieving a flat potential is through
supersymmetry (\susy).  However, the flatness is generically spoiled in
supergravity \cite{Copeland:1994vg},  which must be taken into account
if the inflaton changes by an amount of order the Planck scale or more
(``large-field" inflation). Given the large parameter space of
supergravity theories, this motivates starting the search for a
supersymmetric theory of inflation with small-field inflation, in the
context of a renormalisable theory. 

At the same time,  low energy supersymmetry remains an attractive
theoretical framework in which to understand  the smallness of the
electroweak scale relative to the Planck scale. The Minimal
Supersymmetric Standard Model (\mssm) is the most economical possibility
to combine low energy \susy\ with the phenomenological triumph of the
Standard Model (although the high Higgs mass and the absence of positive
results from the Tevatron and LHC increases the amount of parameter
tuning required).

Indeed, the \mssm\ itself can realise inflation along one of the many
flat directions \cite{Enqvist:2003gh}  with the addition of
non-renormalisable couplings. Inflation takes place near an inflection
point in the potential, where trilinear and soft mass terms are balanced
against each other, although the amount of tuning required
\cite{Allahverdi:2006we} reduces the attractiveness of the scenario. The
tuning can be reduced by extending the \mssm\
\cite{Enqvist:2010vd,Hotchkiss:2011am}.  

The simplest class of renormalisable supersymmetric inflation models is
minimal F-term hybrid inflation, by which we mean the
first supersymmetric model of \reference{Copeland:1994vg}, 
characterised by the superpotential 
\ben
W_I = \lambda_1 \Phi\Phib S  - M^2 S.
\label{e:WI}
\een 
General theoretical considerations of small-field inflation drive one 
towards this model \cite{Dine:2011ws}, which works without a Planck-scale inflaton field, non-renormalisable
operators, or supersymmetry-breaking terms. It invokes an inflaton sector
of (at least) 3 chiral superfields, consisting of the inflaton itself, $S$, 
and two waterfall~\cite{Linde:1993cn}  fields, $\Phi,\Phib$, 
with an optional gauge
superfield.\footnote{The number of chiral superfields can be reduced to
2 without a gauge field, or if they are in a real representation.}
Because the the inflaton field appears linearly in the superpotential, it does not
suffer from the generic supergravity problem of Hubble-scale mass terms
during inflation~\cite{Copeland:1994vg}.

In its standard form, however, F-term hybrid inflation suffers from a
number of problems which reduce its power to fit cosmological data. 
First and foremost is the gravitino problem, which limits the reheat
temperature to be unnaturally small compared with the inflation scale.  
Of less severity is the spectral index problem. If the inflaton 
potential is dominated by the 1-loop radiative corrections, 
F-term hybrid inflation predicts that the  spectral
index of cosmological perturbations $N$ e-foldings before the end of
inflation is $n_s = 1 - 1/N$.  For the canonical 60 e-foldings, this is
more than 1$\sigma$ above the WMAP7 value $n_s = 0.963\pm0.012$. 
Finally, many models generate cosmic strings, and the \cmb\ constraints on
their mass per unit length forces one to very weak inflaton couplings,
where $n_s \to 1$ \cite{Battye:2010hg}. 

There also remains the question of how the inflaton sector is coupled to
the \mssm.  If we restrict ourselves to renormalisable theories
combining minimal U(1)$'$-gauged F-term hybrid inflation with the \mssm,
with no other fields, and preserving all the symmetries, the choices are
limited. The singlet inflaton $S$ can couple in the superpotential only
to the product of the Higgs fields or the square of the right-handed
neutrino fields (which we take to be included the \mssm). If the \mssm\
fields have non-trivial charge assignments under the U(1)$'$ of F-term
inflation, the coupling of $S$ to the neutrinos is forbidden, and its
place taken by  one of the waterfall fields. This
has the nice feature of generating a see-saw mechanism, with the
neutrino masses also controlled by the vev of the waterfall fields. Neutrino
masses are also allowed if the waterfall fields are U(2) triplets,
with SU(2)$_R$ as a subgroup. 

{We will refer to the minimal 
case where the symmetry of the waterfall fields is U(1)$'$ as the 
Minimal Hybrid Inflationary Supersymmetric Standard Model (\mhissm).}
In the model, it is very natural that the gauge singlet inflaton $S$ should be coupled both
to the waterfall fields and to the Higgs fields, which mixes the
standard \mssm\ Higgs flat direction with the hybrid inflation waterfall
direction. If the coupling of the inflaton to the Higgs is smaller than
to the waterfall fields, inflation ends with the development of vevs for
the Higgs multiplets, $h_{1,2}$, breaking the electroweak symmetry.  
Soft terms lift the flat direction, and if certain constraints are
satisfied, the Higgs fields will finally reach the standard vacuum after
a period of thermal inflation, with a reheat temperature of about $10^9$
GeV. This solves the gravitino overclosure problem, and Big Bang
Nucleosynthesis  constraints can be satisfied with massive (O(10) TeV or more) 
or stable gravitinos   
 \cite{Weinberg:1982zq,Nanopoulos:1983up,Khlopov:1984pf,Ellis:1984eq}.
 
We call this second period of accelerated expansion Higgs thermal
inflation. It is a natural consequence of the coupling of the F-term
hybrid inflaton to the Higgs fields, and offers a generic solution to
the gravitino problem. At the same time, a TeV-scale vacuum expectation
value for the inflaton generates an effective $\mu$-term.  The model was
first introduced in  \reference{Hindmarsh:2012wh} in the context of
Anomaly-Mediated Supersymmetry Breaking (\amsb). We termed the version 
of \amsb\ there deployed {\it strictly\/} anomaly mediated supersymmetry 
breaking (\samsb), because D-terms associated with the U(1)$'$
symmetry resolve  the \amsb\ tachyonic slepton problem, without
requiring an additional  explicit source of \sy\ breaking.

In this paper we demonstrate that  the interesting cosmological
consequences, in particular Higgs thermal inflation, are a result of the
structure of the model at the inflation scale, and not  of the
particular supersymmetry-breaking scenario.   We derive the effective
potential for the combination of fields driving thermal inflation, and
the constraints on the soft breaking parameters for a phenomenologically
acceptable ground state, in three popular supersymmetry-breaking
scenarios:  anomaly-mediated (\amsb), gauge-mediated (\gmsb) and  the
constrained minimal supersymmetric standard model (\cmssm). We find that
 the lower reheat temperature following thermal inflation solves  the
gravitino problem in the \cmssm, while in \amsb\ enough gravitinos can
be produced to account for the observed dark matter abundance through 
decays into neutralinos.
 In \gmsb\ we find an upper bound on the gravitino mass of about a TeV,
derived from constraints on \nlsp\ decays during and after Big Bang
Nucleosynthesis (\bbn).

F-term models with Higgs thermal inflation have other important
features.  The spectral index of scalar Cosmic Microwave Background
fluctuations $n_s$ is reduced,  as fewer e-foldings of
F-term inflation are required. In the range of couplings for which the 
1-loop radiative corrections dominate the inflaton potential, 
we find $n_s = 0.976(1)$, where the uncertainty comes from the 
spread of reheat temperatures in that range.
The cosmic string mass per unit length is
greatly reduced by the presence of a Higgs condensate at the string
core, and is rendered independent of the inflaton coupling. Finally,
thermal inflation sweeps away the gravitinos generated at the first
stage of inflation, and any GUT-scale relics such as magnetic monopoles.

There are other models which renormalisably couple F-term hybrid
inflation to the \mssm. $\textrm{F}_{\textrm{D}}$ hybrid inflation 
\cite{Garbrecht:2006ft,Garbrecht:2006az} 
has the same field
content as ours, but the \mssm\ has no U(1)$'$ charges; and it requires 
a Fayet Iliopoulos term. Also
potentially in the class is the B$-$L model of Refs.\
\cite{Buchmuller:2010yy,Buchmuller:2011mw,Buchmuller:2012wn},  
although there is no explicit discussion of the coupling of the inflaton to the
Higgs fields.  In the model of \reference{Dvali:1997uq}  the waterfall
fields are SU(2)$_R$ triplets. The authors identified a flat direction
involving the Higgs, without pursuing its consequences.  The original
F-term inflation model \cite{Copeland:1994vg} had a spontaneously broken
 global U(1) symmetry, and models based on coupling it to the \mssm\
have recently been explored in \cite{Kawasaki:2010gv}, again without the
possibility of Higgs thermal inflation being noticed.  
The same field content can also produce a promising superconformal D-term 
inflation model \cite{Buchmuller:2012ex}.

Further afield, it is also possible to construct renormalisable models
of inflation in the Next-to-Minimal Supersymmetric Standard Model using
soft terms to generate the vacuum energy \cite{BasteroGil:1997vn}. 
Inflation along a flat direction which mixes a singlet with an \mssm\
flat direction has also been investigated recently in
\reference{Nakayama:2012gh}.  In that work, a single stage of inflation
was envisaged, and in order to supply a satisfactory spectral index, the
coupling to the inflaton has to be non-renormalisable.  

The spectral index problem can also be solved with a non-minimal K\"ahler potential \cite{BasteroGil:2006cm}, or 
tuning the inflaton coupling to be small enough that the linear soft term dominates its potential \cite{Rehman:2009nq}.  In this paper we will restrict ourselves to the case where radiative corrections dominate the inflaton potential, and the K\"ahler potential is 
canonical.

\section{Coupling F-term inflation and the \mssm}

Our guiding principle is to couple minimal F-term hybrid inflation and the
\mssm\ (which we take to include 3 families of right-handed neutrinos)
in a renormalisable way, preserving all 
symmetries including supersymmetry (while allowing soft breaking terms in both sectors). 
Hence the superpotential will take the form 
\ben 
W = W_I + W_A + W_X
\label{eq:superpot}
\ee
where $W_I$ is the standard linear F-term hybrid inflation superpotential of 
\eqn{e:WI}, $W_A$ is the \mssm\ Yukawa superpotential 
\ben
W_A = H_2 Q Y_U U + H_1 Q Y_D D + 
H_1 L Y_E E + H_2 L Y_{N} N,
\label{e:WA}
\ee
and $W_X$ is the coupling between the inflaton sector and the \mssm\ superpotential, containing renormalisable terms only. 
We will assume that the U(1)$'$ symmetry of the waterfall fields 
\ben
\Phi \to \Phi' = e^{iq_{\Phi}\th}\Phi, \quad 
\Phib \to \Phib' = e^{iq_{\Phib}\th}\Phib
\een
is gauged.
The inflaton $S$ must be a gauge singlet, and so $q_{\Phib} =
-q_{\Phi}$.  The mass scale $M$ sets the inflation scale and the vevs of
$\Phi$ and $\Phib$.  Given that the inflation scale is of order
$10^{14}$ GeV, the waterfall fields must be
SU(3)$\otimes$SU(2)$\otimes$U(1)$_Y$ singlets.  Note that $W_I$ has a
global U(1) R-symmetry, which forbids the terms $S^2$, $S^3$ and
$\Phi\Phib$.  In order to preserve the flat potential for the inflaton,
we must preserve this symmetry; we will discuss more of 
its implications in a moment.

The form of $W_X$ is now tightly constrained by symmetry and anomaly
cancellation. Possible anomaly-free $\textrm{U(1)}'$ charge
assignments for the \mssm\ fields are shown in Table~1.
%
\begin{table}
\centering
\begin{tabular}{|c|c c c c c c|}
\hline
&$Q$ & $U$ & $D$
& $H_1$ & $H_2$ & $N$ \\ \hline
&& & & & & \\ 
$q$&$-\frac{1}{3}q_L$ & $-q_E-\frac{2}{3}q_L$ & $q_E+\frac{4}{3}q_L$
& $-q_E-q_L$ & $q_E+q_L$ & $-2q_L-q_E$ \\ 
&& & & & & \\ \hline
\end{tabular}
\caption{\label{anomfree}Anomaly free $\textrm{U(1)}$ charges for
lepton doublet, singlet charges $q_L$, $q_E$ respectively.}
\end{table}
The \sm{} gauged $\textrm{U(1)}_Y$ is $q_L= - 1, q_E = 2$.
$\textrm{U(1)}_{B-L}$ is $q_E = -q_L = 1$; in the absence of $N$ this
would have $\textrm{U(1)}^3$ and $\textrm{U(1)}$-gravitational
anomalies. The
diagonal subgroup of SU(2)$_R$ is $q_L= 0, q_E = 1$.  Note that quite
generally $q_{H_1} = -q_{H_2}$,  so we will write $q_{H_2} = -q_{H_1} =
q_H$. We will assume that the \mssm\ fields couple to a U(1)$'$ 
distinct from $\textrm{U(1)}_Y$, i.e. that
$2q_L+q_E\ne0$, and moreover that in the \amsb\ case the 
values of $q_L$ and $q_E$ result in a solution to the 
\amsb\ tachyonic slepton problem~\cite{Hodgson:2007kq}. 
For the resulting sparticle 
spectra in this case, see \reference{Hindmarsh:2012wh}. 
(Note that if the U(1)$'$ does not couple to \mssm\ fields, 
we are driven to $\textrm{F}_{\textrm{D}}$ inflation 
\cite{Garbrecht:2006ft,Garbrecht:2006az}). 
Three SU(3)$\otimes$SU(2)$\otimes$U(1)$_Y$ singlets quadratic in the 
\mssm\ fields are available for $W_X$, 
namely 
$H_1 H_2$, $L H_2$  and $NN$ \cite{Allahverdi:2007zz}.  The U(1)$'$ charge assignments, 
combined with the global R-symmetry, with superfield charges 
\ben
S=2, L=E=N=U=D=Q=1, H_1= H_2 = \Phi = \Phib = 0,
\label{rsym}
\ee
now uniquely specify the coupling term as
\bea
W_X &=& \half\lambda_2 N N \Phi - \lambda_3 S H_1 H_2,
\label{e:WX}
\eea
where we have set $q_{\Phi, \Phib} = \pm (4q_L+2q_E)$ to permit the first 
term. All renormalisable B, L violating interactions and the $NN$ and 
$L H_2$ mass terms 
are forbidden by the U(1)$'$ gauge invariance, and  the superpotential 
\eqn{eq:superpot} contains all renormalisable terms consistent 
with U(1)$'$ and the R-symmetry. Note in particular 
that the R-symmetry forbids the Higgs $\mu$-term $H_1 H_2$. 
Moreover, the R-symmetry forbids the quartic superpotential
terms $QQQL$ and $UUDE$, which are allowed by the $\textrm{U(1)}'$ symmetry,
and give rise to dimension 5 operators capable of
causing proton decay~\cite{Weinberg:1981wj,Sakai:1981pk}. 
In fact the charges in \eqn{rsym}\ 
disallow B-violating operators in the superpotential 
of arbitrary dimension. 

Soft terms break the continuous R-symmetry to the usual R-parity. 
The lightest supersymmetric particle (\lsp) is therefore stable. (From \eqn{rsym}, 
the \lsp\ is a scalar quark or lepton, or a gaugino, or a fermionic Higgs, $S$, $\Phi$ or $\Phib$.) 


To summarise the assumptions which force us to this unique class of
theories, we require a theory with :

\begin{enumerate}

\item The field content of minimal F-term inflation and the \mssm. 

\item The symmetries of minimal F-term inflation and the \mssm.

\item Renormalisable couplings only.

\item An inflaton-sector U(1)$'$ gauge symmetry which is coupled to the \mssm.

\end{enumerate}

Note that if $\Phi$ and $\Phib$ are gauged under a larger symmetry
group, the coupling $NN\Phi$ is not allowed, unless they are triplets of
SU(2)$_R$ and $(N,E)$ are doublets~\cite{Dvali:1997uq}.

The parameters $M,\la_1,\la_3$ are real and positive 
and $\lambda_2$ is a symmetric $3\times 3$ matrix which 
we will take to be 
real and diagonal.  The sign of the $\la_3$ term 
above is chosen because with our conventions, 
in the electroweak vacuum  
\ben
H_1 = \left(\frac{v_1}{\sqrt{2}},0\right)^T\quad\hbox{and}\quad 
H_2 = \left(0, \frac{v_2}{\sqrt{2}}\right)^T
\een
we have $H_1 H_2 \to - \frac{1}{2}v_1 v_2$.

In the following we will denote the SU(3)$\otimes$SU(2)$\otimes$U(1)$_Y$
gauge couplings by $g_3$, $g_2$ and $g_1$, and the U(1)$'$ gauge
coupling by $g'$. The normalisation of the U(1)$_Y$ gauge
coupling  corresponds to the usual \sm\ convention, not that appropriate
for $\textrm{SU(5)}$ unification.  We will denote the soft parameters
for the gaugino masses $M_a$, for a cubic  interaction with Yukawa
coupling $\la$   $h_\la$,  and for a mass term  $\phi^*\phi$ (where
$\phi$ denotes a scalar field), $m^2_\phi$.  For the one mass term of
the form $\phi^2$ in the \mssm\ ($H_1 H_2$) we will use $m_3^2$.

\section{The Higgs potential and its extrema}
\label{s:HigPot}

In this section we explore the important extrema of the Higgs potential,
and demonstrate that there is a 1-parameter family of supersymmetric
ground states with non-zero vevs for $\phi, \phib$ and $h_{1,2}$ before
supersymmetry-breaking is taken into account. We will assume that $M$,
the scale of inflation and $\textrm{U(1)}'$ symmetry-breaking, is much
larger than the  scale of \sy-breaking. 

The existence of the one-parameter family (before thermal effects and soft terms are taken into
account), is demonstrated as follows. The minimum of the scalar potential is determined by the
requirement that both the F- and D-terms vanish. The vanishing of the
D-terms ensures that $|{\phi}| = |{\phib}|$, $|h_1| = |h_2|$ and
$h_1^\dag h_2 = 0$, while the vanishing of the F-term is assured by
$\lambda_1\phi\phib - \lambda_3 h_1 h_2 = M^2$.  The minimum can
therefore be parametrised by an SU(2) gauge transformation and angles
$\chi,\varphi$ defined by 
\bea
\vev{h_1}& \simeq& i\si_2\vev{h_2}^* \simeq (\frac{M}{\sqrt\la_3}\cos\chi,0), 
\nn
\vev{\phi} &\simeq & \vev{{\phib}^*} 
\simeq \frac{M}{\sqrt{\la_1}}\sin\chi e^{i\varphi}.
\label{e:ChiDef}
\eea
The $\varphi$ angle can always be removed by a U(1)$''$ gauge
transformation (where the residual symmetry unbroken by the Higgs vevs alone is 
$\textrm{U(1)}_\textrm{em}\times\textrm{U(1)}''$),
so the physical flat direction just maps out the
interval $0 \le \chi \le \pi/2$. At the special point $\chi = 0$ the
U(1)$''$ symmetry is restored, and at $\chi=\pi/2$ the
$\textrm{SU(2)}\otimes\textrm{U(1)}_Y$ is restored. Away from these special points only
U(1)$_\textrm{em}$ is unbroken.

The degenerate minima have been noted before~\cite{Dvali:1997uq} 
in a model with gauge group $\textrm{SU(3)} \otimes \textrm{SU(2)}_L \otimes \textrm{SU(2)}_R \otimes \textrm{U(1)}_{B-L}$. However, the important cosmological consequences which follow was first explored in \reference{Hindmarsh:2012wh}.

Let us first consider the limiting cases where either $h_{1,2}$ or $\phi, \phib$ vanish.
 
\subsection{The $\phi, \phib, s$ extremum ($\phi$-vacuum)}
\label{sec:phiphib}
In the $\phi, \phib, s$ subspace 
(lower case fields denote the scalar component of the superfields) 
the scalar potential (including soft supersymmetry-breaking terms) is:
\bea
V &=& \lambda_1^2 (|\phi s|^2 + |\phib s|^2) + |\lambda_1\phi\phib-M^2|^2
+\frak{1}{2}q_{\Phi}^2 g'^2\left(|\phi|^2 - |\phib|^2\right)^2\nn
&+&m_{\phi}^2 |\phi|^2 + m_{\phib}^2 |\phib|^2 +m_s^2|s|^2 
+\rho M^2 m_{\frac{3}{2}} (s + s^*)\nn
&+& h_{\lambda_1}\phi\phib s + c.c..
\label{eq:treepot}
\eea
We will assume that the term linear in $s$ is small enough not to be important for inflation 
(and quantify this smallness in 
Section \ref{s:Inf}).  In \amsb\ there are arguments~\cite{Jack:2001ew} to show that, without 
a quadratic term $S^2$ in the superpotential, the only \rg\ invariant solution for $\rho$ is 
$\rho = 0$.  

Let us establish the minimum in this subspace, under the assumption that
$\mgrav \ll M$.   We shall call this the $\phi$-vacuum. With the
notation  
$\vev\phi = v_{\phi}/\sqrt{2}$,
$\vev\phib = v_{\phib}/\sqrt{2}$ and 
$\vev s = v_s/\sqrt{2}$, we find 
\bea
v_{\phi}\left[m_{\phi}^2+\half\la_1^2v_s^2
+\half g^2 q_{\Phi}^2(v_{\phi}^2-v_{\phib}^2)\right]
+v_{\phib}\left[\la_1\left(\half\lambda_1 v_{\phi} v_{\phib}-M^2\right)
+\frac{h_{\lambda_1}}{\sqrt{2}}v_s\right]&=&0,
\label{eq:vphi}\\
v_{\phib}\left[m_{\phib}^2+\half\la_1^2v_s^2
-\half g^2 q_{\Phi}^2(v_{\phi}^2-v_{\phib}^2)\right]
+v_{\phi}\left[\la_1\left(\half\lambda_1 v_{\phi} v_{\phib}-M^2\right)
+\frac{h_{\lambda_1}}{\sqrt{2}}v_s\right]&=&0,\label{eq:vphib}\\
v_s\left[m_s^2+\half\la_1^2(v_{\phi}^2+v_{\phib}^2)\right]
+\frac{h_{\lambda_1}}{\sqrt{2}}v_{\phi}v_{\phib}+\sqrt{2}\rho M^2 m_{\frac{3}{2}} &=&0.
\label{eq:vs}
\eea
From \eqns{eq:vphi}{eq:vphib} we find 
\bea
\la_1\left(\half\lambda_1 v_{\phi} v_{\phib}-M^2\right)
&=& -\frac{v_{\phi}v_{\phib}}{v_{\phi}^2+v_{\phib}^2}
\left[m_{\phi}^2+m_{\phib}^2+\la_1^2 v_s^2\right] 
-\frac{h_{\lambda_1}}{\sqrt{2}}v_s,\label{eq:Mterm}\\
\half {g'}^2 q_{\Phi}^2(v_{\phi}^2-v_{\phib}^2)
&=& \frac{v_{\phib}^2m_{\phib}^2-v_{\phi}^2m_{\phi}^2+ 
(v_{\phib}^2-v_{\phi}^2)\half\la_1^2v_s^2}{v_{\phi}^2+v_{\phib}^2}.\label{eq:gterm}
\eea
Then from \eqns{eq:Mterm}{eq:gterm}, to leading order in an expansion in $\mgrav/M$ we have  
\ben
v_{\phi}^2 \simeq v_{\phib}^2 \simeq \frac{2}{\la_1}M^2,
\label{eq:vsize}
\een
and from \eqn{eq:vs} that $v_{s}$ is $O(m_{\frac{3}{2}})$. 
It follows from \eqn{eq:gterm} that 
\ben
v_{\phi}^2-v_{\phib}^2 = \frac{m_{\phib}^2-m_{\phi}^2}{{g'}^2q_{\Phi}^2} 
+ O(m_{\frac{3}{2}}^4/M^2),
\label{eq:vdiff}
\een
and from \eqn{eq:vs} that
\ben
v_s = -\frac{h_{\lambda_1}}{\sqrt{2}\la_1^2}-\frac{m_{\frac{3}{2}}\rho}{\sqrt{2}\la_1}
+ O(m_{\frac{3}{2}}^2/M). 
\label{eq:vsans}
\een
From now on we neglect $\rho$, assuming that 
\ben
|\rho| \lesssim \left|\frac{h_{\lambda_1}}{\la_1\mgrav}\right|.
\een
Substituting back from \eqns{eq:vsize}{eq:vsans} into \eqn{eq:treepot}, we obtain to leading 
order 
\ben
V_{\phi} 
= \frac{1}{\la_1}M^2\left(m_{\phi}^2 +m_{\phib}^2 -\frac{h_{\la_1}^2}{2\la_1^2}\right)
\label{eq:phipot}
\een
and from \eqn{eq:vsans} a Higgs $\mu$-term 
\ben
\mu_h = \frac{\la_3 h_{\la_1}}{2\la_1^2},
\label{eq:muans}
\een
naturally of the same order as the \sy-breaking scale.  

The theory is approximately supersymmetric at the scale $M$,  so the
$\textrm{U(1)}'$ gauge boson, the Higgs boson, the gaugino  and one
combination of $\psi_{\phi,\phib}$ form a massive supermultiplet with
mass $m \sim g'\sqrt{v_{\phi}^2 + v_{\phib}^2}$, while the remaining
combination of $\phi$ and $\phib$ and the other combination of
$\psi_{\phi,\phib}$ form a massive chiral supermultiplet, with mass  $m
\sim \lambda_1\sqrt{v_{\phi}^2 + v_{\phib}^2}$.\footnote{A detailed
explanation of the symmetry-breaking is contained  in
\reference{Buchmuller:2012wn}.}

The large vev for $\phi$ generates inflation-scale masses for the $N$
triplet, thus naturally implementing the see-saw mechanism.

\subsection{The $h_{1,2}, s$ extremum ($h$-vacuum)}.
In the $h_{1,2}, s$ subspace,  the scalar potential is 
\bea
V &=& \lambda_3^2 (|h_1 s|^2 + |h_2 s|^2) + |\lambda_3 h_1 h_2-M^2|^2
+\frak{1}{2}g'^2 q_{H}^2\left(|h_1|^2 - |h_2|^2\right)^2\nn
&+& \frak{1}{8}g_1^2 (h_1^{\dagger}h_1
 -h_2^{\dagger}h_2)^2 + \frak{1}{8}g_2^2\sum_{a}(h_1^{\dagger}\sigma^a h_1
+h_2^{\dagger}\sigma^a h_2)^2\nn
&+&m_{h_1}^2 |h_1|^2 + m_{h_2}^2 |h_2|^2 +m_s^2|s|^2 
+\rho M^2 m_{\frac{3}{2}} (s + s^*)\nn
&+& h_{\lambda_3}h_1 h_2 s + c.c..
\label{eq:treepoth}
\eea
Note that we assume there is no $h_1 h_2$ mass term; its absence follows 
from the absence of the corresponding term in the superpotential (which is 
forbidden by the R-symmetry) when the source of supersymmetry breaking 
can be represented by a non-zero vev for a spurion (or conformal compensator) 
field. 

The structure is similar to \eqn{eq:treepot}, with the addition of $\textrm{SU(2)}$ and $U(1)_Y$ D-terms. 
Without loss of generality the $\textrm{SU(2)}$ D-term vanishes with the choice 
$h_1 = (v_1/\sqrt{2},0)$ and $h_2 = (0,v_2/\sqrt{2})$, and $v_1 = v_2$. 
The values of the fields at the minimum (which we term the $h$-vacuum) and the 
value of the potential at this extremum can 
then be recovered from the result of the previous section 
with the replacement $\la_1 \to \la_3$), leading to a potential energy density 
\ben
V_h = \frac{M^2}{\la_3}
\left(m_{h_1}^2+m_{h_2}^2-\frac{h_{\la_3}^2}{2\la_3^2}\right).
\label{eq:hpot}
\een

\subsection{Potential along the $\phi$, $\phib$, $h_1$, $h_2$ flat direction}

As we outlined at the beginning of the section, the supersymmetric minima are parametrised by an angle $\chi$, defined in (\ref{e:ChiDef}).  Soft terms lift this degeneracy, and the leading terms in the effective potential for $\chi$  can be found in an expansion in $m_\frac32^2/M^2$. After solving for $s$, it is found that 
\ben
V(\chi) \simeq -\frac{M^2}{2}\frac{\left( \tilh_{\la_1}\sin^2\chi + \tilh_{\la_3}\cos^2\chi\right)^2}{\la_1\sin^2\chi +{\la_3}\cos^2\chi}
+ M^2 \left( \frac{\barm_\phi^2}{\la_1}\sin^2\chi + \frac{\barm_h^2}{\la_3}\cos^2\chi \right),
\een
where we have defined 
\ben
\tilh_{\la_1} = \frac{h_{\la_1}}{\la_1}, \quad \tilh_{\la_3}=\frac{h_{\la_3}}{\la_3}, \quad \barm_\phi^2 = m_\phi^2 + m_{\bar\phi}^2, \quad \barm_h^2 = m_{h_1}^2 + m_{h_2}^2.
\een

\section{Supersymmetry-breaking and the true minimum}
\label{s:TruMin}

In this section we investigate under which conditions the
phenomenologically acceptable large-$\phi$ solution  is the true
minimum, in three popular supersymmetry-breaking scenarios.  Hence we
are looking for constraints on the soft supersymmetry-breaking
parameters such that
\bea
V_h - V_\phi &=& M^2\left(  \frac{\tilh_{\la_1}^2}{2\la_1}- \frac{\tilh_{\la_3}^2}{2\la_3} - \frac{\barm_\phi^2}{\la_1} + \frac{\barm_h^2}{\la_3} \right) > 0, 
\label{e:VCon}\\
V''(\pi/2) &=& \frac{2M^2}{\la_1}\left[-\frac{\tilh_{\la_1}^2}{2} 
\left(2 \frac{\tilh_{\la_3}}{\tilh_{\la_1}} - \frac{\la_3}{\la_1} - 1\right)  + \barm_h^2\frac{\la_1}{\la_3} - \barm_\phi^2\right] > 0.
\label{e:VppCon}
\eea
We will also check that the false vacuum at $\chi = 0$ is a local
maximum, from the sign of $V''(0)$, which can be recovered from
$V''(\pi/2)$ by the replacements $1 \leftrightarrow 3$ and $\barm^2_\phi
\leftrightarrow \barm^2_h$. 
A metastable false vacuum, 
as we will demonstrate in Section \ref{s:HigTheInfGra},  
would lead to the universe remaining trapped in an inflating phase.

We assume that the U(1)$'$ symmetry is broken by a vev of order $v' \sim
M/\sqrt{\la_{1,3}}$ , and evaluate the soft terms at this scale, rather
than running down to the electroweak scale.  This is the appropriate 
renormalisation scale to investigate a potential with vevs of order
$v'$, whose important radiative corrections are from particles of mass
of order $g'v'$ and $M$.  Note that in inflation models, with inflaton
couplings $\la_1$ and $\la_3$ are generally small, and so the U(1)$'$
gauge boson mass $m_A = g'\sqrt{v_{\phi}^2 + v_{\phib}^2}$ is much
greater than $M$, unless $g'$ is also small. 

\subsection{Anomaly-mediated supersymmetry-breaking }
\label{s:TruMinAMSB}

With anomaly mediation, the soft breaking parameters  
take the generic renormalisation group invariant form
\bea
M_a & = & m_{\frac{3}{2}} \beta_{g_a}/{g_a},\label{eq:AD1}\\
h_{U,D,E,N} & = & -m_{\frac{3}{2}}\beta_{Y_{U,D,E,N}},\label{eq:AD2}\\
(m^2)^i{}_j &=& \frac{1}{2}m_{\frac{3}{2}}^2\mu\frac{d}{d\mu}\gamma^i{}_j
+kY'_i\delta^i{}_j,\label{eq:AD3b} \\
m_3^2 & = & \kappa m_{\frac{3}{2}} \mu_h - m_{\frac{3}{2}} \beta_{\mu_h}.
\label{eq:AD4}
\eea
Here $\mu$ is the renormalisation scale,  and $m_{\frac{3}{2}}$ is the
gravitino mass;  $\beta_{g_a}$ are the gauge $\beta$-functions and
$\gamma$ is the  chiral supermultiplet anomalous dimension matrix. 
$Y_{U,D,E,N}$ are the $3 \times 3$ Yukawa matrices, $\mu_h$ is the
superpotential Higgs $\mu$-term,  $\ka$ and $k$ are  constants, and
$Y'_i$ are charges corresponding to the $\textrm{U(1)}'$ symmetry. 

In the \mssm, $\kappa$ is an arbitrary parameter, which in
practice is fixed by minimising the Higgs potential at the electroweak
scale.  The parameter $k$ is generated by the breaking of the
$\textrm{U(1)}'$ symmetry at a large scale, and forms the  basis of the
solution to the tachyonic slepton problem  within the framework
of \amsb, as explained in \cite{Hindmarsh:2012wh}, whence the name
strictly anomaly-mediated supersymmetry-breaking (\samsb) originates. 

The Higgs $\mu$-term,   $\mu_h$, is generated by the the vev of the
inflation $s$, which in turn is triggered by the U(1)$'$
symmetry-breaking.  Hence the parameter $k$, and the equation for
$m_3^2$, are relevant only below the $\textrm{U(1)}'$ symmetry-breaking
scale $v'$.

As a first approximation, we will assume that the $g'$ terms dominate
throughout, as $q_H$ and $q_{\Phi}$ are generally large, in which  case
the $h_{\lambda_1}$ and $h_{\lambda_3}$ trilinear  soft terms are given
from  \eqn{eq:AD2} as:
\bea
h_{\lambda_1} &\simeq& m_{\frac{3}{2}}\frac{\la_1}{16\pi^2}\left(4q_{\Phi}^2{g'}^2 \right), 
\label{eq:hlaone}\\
h_{\lambda_3} &\simeq& m_{\frac{3}{2}}\frac{\la_3}{16\pi^2}(4q_{H}^2{g'}^2), 
\eea
while the mass soft terms are
\bea
m_\phi^2 & \simeq &  -\mgrav^2 \frac{1}{32\pi^2} \mu 
\frac{d}{d\mu}\left( 2{g'}^2q_{\Phi}^2\right),\\
m_{\phib}^2 & \simeq & -\mgrav^2 \frac{1}{32\pi^2} \mu 
\frac{d}{d\mu}\left( 2{g'}^2q_{\Phi}^2\right), \\
m_{h_1}^2 & \simeq &  - \mgrav^2 \frac{1}{32\pi^2} \mu 
\frac{d}{d\mu}\left( 2{g'}^2q_{H}^2\right), \\
m_{h_2}^2 & \simeq & - \mgrav^2 \frac{1}{32\pi^2} \mu 
\frac{d}{d\mu}\left( 2{g'}^2q_H^2\right).
\eea
The  one loop $g'$ $\beta$-function is
\ben
\beta_{g'} = Q \frac{{g'}^3}{16\pi^2}
\een
where 
\bea
Q &=& n_G (\frak{40}{3}q_L^2+8 q_E^2+16q_E q_L)+36q_L^2+40q_E q_L+12q_E^2\nn
&=& 76q_L^2+36q_E^2+88q_E q_L
\label{e:Qeqn}
\eea
 {for $n_G = 3$}.
Hence
\bea
m_\phi^2 \simeq m_{\phib}^2 & \simeq &  
-2 \mgrav^2 \left(\frac{{g'}^2}{16\pi^2}\right)^2 q_{\Phi}^2Q,  \\
m_{h_1}^2 \simeq m_{h_2}^2  & \simeq &  -2 \mgrav^2 \left(\frac{{g'}^2}{16\pi^2}\right)^2 q_H^2Q . 
\eea
Thus the difference in the energy densities between the two vacua is, in this approximation, 
\ben
V_{h} - V_\phi \simeq {M^2}\left(\frac{m_{\frac{3}{2}}{g'}^2}{16\pi^2}\right)^2
\left[\frac{4Q q_{\Phi}^2 + 8q_{\Phi}^4}{\la_1} - \frac{4Q q_{H}^2 + 8q_{H}^4}{\la_3}\right].
\label{eq:VdiffAMSB}
\een 
The coefficient $Q$ is in general large, and larger than both
$q^2_\Phi$ and $q^2_H$, so the condition for $V_\phi$ to be the true minimum may be written 
\ben
\label{e:EWCon}
\frac{\la_3}{\la_1}\gtrsim \left(\frac{q_{H}}{q_{\Phi}}\right)^2.
\een
It is not hard to check from \eqn{e:VppCon}) 
that under the same assumptions, the $\phi$-vacuum is a 
minimum and the $h$-vacuum is a maximum. 
Hence no further constraints on the parameters are generated.

In the next section we will see that if $\la_3 > \la_1$, then inflation ends 
with $\phi,\phib$ developing non-zero vevs, whereas if $\la_3 < \la_1$ 
it is $\vev{h_{1,2}}$ which become non-zero; this statement is independent 
of the nature of the soft breaking terms.  Now is easy to show 
that $ \left(\frac{q_{H}}{q_{\Phi}}\right)^2 < 1$ unless 
\ben
-\frac{3}{5} \leq \frac{q_L}{q_E} \leq -\frac{1}{3}.
\label{eq:LErange}
\een
However, the domain defined by \eqn{eq:LErange}\ does not 
permit a satisfactory electroweak vacuum in the \amsb\ case~\cite{Hodgson:2007kq}.
For example, for the specific choice $q_L = 0$, which {\it can\/} lead to an 
acceptable electro-weak vacuum~\cite{Hindmarsh:2012wh}, 
the condition $V_h > V_{\phi}$ becomes (from \eqn{eq:VdiffAMSB})
\ben
\label{e:EWConb}
\frac{\la_3}{\la_1}\gtrsim \frac{19}{88}.
\een
or $\la_1 \lesssim 4\la_3$ from the approximation \eqn{e:EWCon}.

We see, therefore, that there will generally be a domain 
\ben
 \la_1\left(\frac{q_{H}}{q_{\Phi}}\right)^2 \lesssim \la_3 < \la_1
\label{eq:la3range}
\een
such that the universe exits to the false high Higgs vev $h$-vacuum, evolving 
subsequently to the true vacuum as we shall describe later. 

In the Appendix we include a more accurate computation of the vacuum energy difference, 
taking into account the \sm\
gauge couplings and the top Yukawa coupling.

\subsection{Gauge-mediated supersymmetry-breaking }

In the \gmsb\ framework (see e.g. \cite{Giudice:1998bp}),
supersymmetry-breaking is communicated by a set of messenger fields $C$
which have \sm\ gauge charges in a vector-like representation, which
should be complete GUT multiplets if gauge unification is to be
preserved. The messenger fields are supposed to have a large mass, given
by the vev of the scalar component of a chiral superfield $X$, which also has a non-zero F-term
$F_X$, the source of the supersymmetry breaking. Although there are many
possibly choices for the field  representations of the messenger fields,
we can adapt the simple model described in \cite{Giudice:1998bp} to
study our model.

We introduce the following superpotential for the extra fields  
\ben
W_{gm} = \la_4 S C\bar C + \la_5X C\bar C,
\een
assuming that some extra dynamics at a higher scale gives both the 
scalar component of $X$ and $F_X$ a vev. We will assume that $\vev X\gg M$. 
Radiative corrections from the messenger particles then induce masses for the gauginos at one loop, 
\ben
\label{e:MguaginoGMSB}
M_a = \frac{g^2_a}{16\pi^2}\La_\text{g},
\een
where $\La_\text{g} = N_{\rm mi}\vev{F_X}/M_X$, $M_X = \la_5\vev{X}$, and $N_{\rm mi}$ is the messenger index, equal to twice the sum of the Dynkin indices of the messenger fields.  Scalars acquire masses from 2-loop corrections of
\ben
m^2_i = 2\La_\text{s}^2\sum_{a}\left(\frac{g^2_a}{16\pi^2}\right)^2 C_a (i),
\een
where $\La_\text{s}^2 = N_{\rm mi}(\vev{F_X}/M_X)^2$, $C_a(i)$ is the quadratic Casimir associated with the $a$th gauge group for the $i$th scalar, and the sum over $a$ 
includes the four gauge couplings $g_{1\to 3}, g'$. 

Trilinear terms are also induced at 2 loops, and so are of order
$\La_g(\al_a/4\pi)^2$. They are small compared with the gaugino masses,
and it is a reasonable approximation to take them to vanish at the
messenger scale $M_X$.
We assume that $\La_{g,s}$ are of the correct order of magnitude
for supersymmetry-breaking.

We thus have
\bea
m_{\phi}^2 &=& m_{\phib}^2 = 
2\La_\text{s}^2 \left(\frac{{g'}^2}{16\pi^2}\right)^2q_{\Phi}^2, \nn
m_{h_1}^2 &=& m_{h_2}^2 = 2\La_\text{s}^2\left[ 
\frac{3}{4} \left(  \frac{g^2_2}{16\pi^2} \right)^2 
+ \frac{1}{4} \left(  \frac{g^2_1}{16\pi^2} \right)^2 
+ \left(\frac{{g'}^2}{16\pi^2}\right)^2 q_H^2 
\right], \nn
\frac{h_{\la_1}}{\la_1} &=& \frac{h_{\la_3}}{\la_3} = 0.
\eea
Thus the difference between the vacuum energies  is 
\bea
V_h - V_{\phi} &=& M^2 \left( \frac{m_{h_1}^2 + m_{h_2}^2}{\la_3} 
- \frac{m_{\phi}^2 + m_{\phib}^2}{\la_1}\right)\nn
&=&\frac{2\La_\text{s}^2 M^2}{(16\pi^2)^2}
\left[\left(\frac{3}{2}g_2^4 + \frac{1}{2} g_1^4 + 2 q_H^2 {g'}^4\right)\frac{1}{\la_3} 
 - \frac{2 q_{\Phi}^2{g'}^4}{\la_1}\right],
\label{eq:VdiffGMSB}
\eea
so that, if we assume dominance of the $g'$ terms,  
the condition that $V_{\phi}< V_h $ becomes 
\ben
\label{e:gmsb}
\frac{\la_3}{\la_1} \lesssim \left(\frac{q_{H}}{q_{\Phi}}\right)^2  .
\een
This is precisely the opposite condition to that in \amsb, \eqn{e:EWCon}.  As in \amsb, the condition that $V_{\phi}< V_h $ is sufficient to ensure that $V_{\phi}$ is a minimum and $V_h$ a maximum.

Now in \gmsb, we do not have the constraint on the domain $(q_L,q_E)$ that we described in the \amsb\ case. 
Inflation will end in the Higgs phase unless 
\ben
\left(\frac{q_{H}}{q_{\Phi}}\right)^2 > 1\quad \hbox{and} \quad  
1 < \frac{\la_3}{\la_1} < \left(\frac{q_{H}}{q_{\Phi}}\right)^2,
\een
in which case it ends directly in the true $\phi$-vacuum.

\subsection{Constrained minimal supersymmetric standard model}

At the high scale 
we will have the \cmssm\  pattern of soft breaking parameters,  
\bea 
m_{\phi}^2 &=& m_{\phib}^2 = m_{h_1}^2 = m_{h_2}^2 = m_0^2,\nn
\frac{h_{\la_1}}{\la_1} &=& \frac{h_{\la_3}}{\la_3} = A
\eea
and hence
\ben
V_h - V_{\phi} 
=  M^2 (2m_0^2 -A^2/2)\left[\frac{1}{\la_3} -\frac{1}{\la_1}\right].
\label{eq:VdiffmSUGRA}
\een
Hence if  $\la_3 < \la_1$ (so that inflation ends in the $h$-vacuum) 
then for $V_h > V_{\phi}$ we require   
\ben
 2m_0^2 > A^2/2.
\een
It is easy to check from \eqn{e:VppCon} that this is again a 
sufficient condition that  $V''(\pi/2)$ be positive.
On the other hand, there is then a range 
\ben
\frac{A^2}{2} < 2m_0^2 < \frac{\la_1}{\la_3}\frac{A^2}{2}
\een
for which the $h$-vacuum is {\it also\/} a local minimum.   
We will see that this scenario is not consistent with 
a graceful exit from Higgs thermal inflation, and hence for a cosmologically acceptable potential, we must demand
\ben
2m_0^2 > \frac{\la_1}{\la_3}\frac{A^2}{2}.
\label{eq:sugrahigg}
\een

\section{Inflation and reheating}

\label{s:Inf}

\subsection{F-term inflation}

We assume that the vevs of \mssm\ fields apart from the Higgs are negligible, in which 
case the relevant tree potential is
\bea
V_{\rm tree} &=& |\lambda_1\phi\phib - \lambda_3 h_1 h_2 - M^2|^2 + 
\left[\lambda_1^2 (|\phi|^2 + |\phib|^2) 
+ \lambda^2_3 (|h_1 |^2 +|h_2 |^2)\right] | s |^2
\nn 
&+&\frac{1}{2}g'^2\left(q_{\Phi}(\phi^*\phi -\phib^*\phib)
+ q_{H} (h_1^{\dagger}h_1- h_2^{\dagger}h_2)\right)^2\nn
&+& \frak{1}{8}g_2^2 \sum_{a}(h_1^{\dagger}\sigma^a h_1 
+h_2^{\dagger}\sigma^a h_2 )^2 + \frak{1}{8}g_1^2 ( h_1^{\dagger}h_1
 - h_2^{\dagger}h_2)^2\nn
&+& V_\mathrm{soft}.
\label{eq:treepotb}
\eea
The soft terms in $V_\mathrm{soft}$ are those appearing
in \eqns{eq:treepot}{eq:treepoth}, and are all suppressed  by at least
one power of $m_{\frac{3}{2}}$. The most important soft term for inflation is 
one linear in $s$, 
the effect of which we assume is small compared with the radiative correction.  
We will see in \eqn{e:RhoCon} that this implies tuning below O(1) only if the couplings 
$\la_{1,3}$ are very small.
We also assume that the higher order terms in the K\"ahler potential 
do not contribute significantly.  

At large $s$, and with all other fields vanishing, the potential is approximately 
\ben
V = M^4 + \Delta V_1,
\ee
where $\Delta V_1$ represents the one-loop corrections, which dominate the soft terms.  
As $S$ is coupled only to $\Phi$, $\Phib$ and $H_{1,2}$, 
the contribution to the one-loop scalar potential is \cite{Basboll:2011mh}
\bea
\Delta V_1 
&=&\frac{1}{32\pi^2}\left[(\lambda_1^2 s^2+\lambda_1 M^2)^2\ln 
\left(\frac{\la_1^2s^2+\lambda_1 M^2}{\mu^2}\right)
+ (\lambda_1^2 s^2-\lambda_1 M^2)^2\ln 
\left(\frac{\la_1^2s^2-\lambda_1 M^2}{\mu^2}\right) \right.\nn 
&+& 2(\lambda_3^2 s^2+\lambda_3 M^2)^2\ln 
\left(\frac{\la_3^2s^2+\lambda_3 M^2}{\mu^2}\right)
+ 2(\lambda_3^2 s^2-\lambda_3 M^2)^2\ln 
\left(\frac{\la_3^2s^2-\lambda_3 M^2}{\mu^2}\right)\nn
&-&\left. 2\lambda_1^4 s^4\ln\left(\frac{\la_1^2s^2}{\mu^2}\right)
 -4\lambda_3^4 s^4\ln\left(\frac{\la_3^2s^2}{\mu^2}\right)
\right].
\label{e:1-loopDeV}
\eea
For large $s$ (meaning $\la_{1,3} s^2 \gg M^2$) the potential can be written as  
\ben 
V(s) \simeq M^4 \left[1+ \al\ln
\frac{2s^2}{s_c^2}\right], 
\label{e:VsApp}
\een 
where an O($\al$) correction to $M^4$ has been dropped, and 
\ben 
\alpha = \frac{\lambda^2}{16\pi^2}, \quad \la = 
\sqrt{\lambda_1^2+ 2\lambda_3^2}, \quad  s_c^2 = M^2/\lambda.
\label{eq:newalpha}
\een
We will neglect supergravity contributions in the potential, which will
require a small coupling $c$ of the quartic term $c|s|^4/\mpl^2$ in the
Kahler potential, and impose a constraint \cite{Linde:1997sj}
\ben
\label{e:LamUppBou}
\la \lesssim 0.06.
\een
There are also potentially important contributions from the linear soft term $\rho M^2 m_\frac32 s + \textrm{c.c.}$.  These are negligible provided 
\ben
\label{e:RhoCon}
\rho \ll \frac{\la^3}{16\pi^2} \frac{s_c}{m_\frac32}.
\een
We will shortly see that $s_c \sim 10^{16}\; \text{GeV}$, so assuming $\mgrav \sim  10^5\;\text{GeV}$, a soft term with $\rho \sim 1$ is negligible provided 
\ben
\label{e:LamLowBou}
10^{-3} \lesssim \la.
\een 
Henceforth we will assume that the K\"ahler potential is canonical and that $\la$ is 
in the range given by \eqns{e:LamUppBou}{e:LamLowBou}.  We note, however, that interesting consequences for the spectral index flow from a non-canonical K\"ahler potential \cite{BasteroGil:2006cm} and from couplings small enough for the soft term to contribute \cite{Rehman:2009nq}.

\subsection{Perturbation amplitudes}
The scalar and tensor power spectra $\powspec_s$, $\powspec_t$ and the scalar spectral index $n_s$ generated on a scale $k$ equal to the co-moving Hubble scale $aH$ at  $N_k$ e-foldings before the end of inflation are given by the standard formulae (see e.g. \cite{Lyth:2009zz}), 
\bea 
\powspec_s(k) & \simeq & \frac{1}{24\pi^2} \frac{2N_k}{\al}
\left(\frac{M}{m_p}\right)^{4} = \frac{4N_k}{3}  \left(\frac{s_c}{m_p}\right)^{4}, \\ 
\powspec_t(k) &\simeq & 
\frac{1}{6\pi^2}\left(\frac{M}{m_p}\right)^{4} = \frac83 \al \left(\frac{s_c}{m_p}\right)^{4} , \\ 
n_s & \simeq & \left(1 -
\frac{1}{N_k}\right). 
\eea 
The WMAP7 best-fit values for $\powspec_s(k_0)$ and $n_s$ at a pivot scale $k = k_0 = 0.002\;h\textrm{Mpc}^{-1}$ 
in the standard $\La$CDM model are \cite{Komatsu:2010fb}
\ben\label{eq:WMAP7ps}
\powspec_s(k_0) = (2.43\pm0.11) \times 10^{-9}, \quad n_s = 0.963\pm0.012 (68\% \textrm{CL}).
\een
From this data we infer that 
\ben
\label{e:ScLim}
\frac{s_c}{\mpl} \simeq 2.9 \times 10^{-3} \left(\frac{27}{N_{k_0}}\right)^\frac14, \quad N_{k_0} = 27^{+13}_{-7}, 
\een
showing approximately a 2$\si$ discrepancy with the standard Hot Big Bang result
$N_{k_0} \simeq 58 + \ln(T_\mathrm{rh}/10^{15}\;\mathrm{GeV})$ (assuming only \mssm\ degrees of freedom at  
$T_\mathrm{rh}$).  
We will see shortly that the reheat temperature lies in a range around $10^{14}$ GeV, and 
in Section \ref{s:HigTheInfGra} that there are $N_\th \simeq 15$ e-foldings of thermal inflation at a lower scale. 
Therefore one can estimate $N_{\text{Fti}} \simeq 42(1)$ e-foldings of F-term inflation while the pivot scale $k_0$ is outside the horizon, 
where the uncertainty comes from the range of reheat temperatures, given in \eqn{e:Trh1Ran}.  
The scalar spectral index is thereby reduced to 
\ben
n_s  \simeq  \left(1 -
\frac{1}{N_{\text{Fti}}}\right) \simeq 0.976(1).
\een
Lower values of the spectral index are possible if $\la$ drops below the limit (\ref{e:LamLowBou}) and the linear soft term comes into play \cite{Rehman:2009nq}.

\subsection{End of inflation and reheating}

F-term inflation ends when one set of scalar fields becomes unstable. 
If $\la_3 > \la_1$, the $\phi$, $\phib$ pair become unstable first, and 
inflation ends at the critical value 
$
 s^2_{c1} = M^2/\la_1.
$
The fields $\phi$, $\phib$ gain vevs and the universe makes a transition to the $\textrm{U(1)}'$-broken phase described by 
\eqn{eq:vphi}-\eqn{eq:vs}. 
On the other hand, if $\lambda_3 < \lambda_1$, the Higgs fields become unstable first, the critical value of $s$ is $s_{c3}^2 = M^2/\lambda_3$, and the universe makes a transition to a phase where $h_1$ and $h_2$ develop vevs of order the unification scale rather than $\phi, \phib$. In this phase the $SU(2)_L$ symmetry is broken.

At first sight, this would appear to rule out the model with $\lambda_3 < \lambda_1$. 
However, provided the correct (small Higgs vev)
vacuum has the lowest energy density   
at zero temperature, the universe can seek the true vacuum when thermal corrections become sub-dominant.
We will establish in Section \ref{s:HigTheInfGra} that the evolution to the true ground state proceeds by a period of inflation.

Assuming that $\lambda_3 < \lambda_1$, inflation exits to the $h$-vacuum,
with symmetry-breaking  
\ben
\textrm{SU(2)}\otimes\textrm{U(1)}_Y\otimes \textrm{U(1)}' \quad \to \quad 
\textrm{U(1)}_\textrm{em}\otimes\textrm{U(1)}''.
\een
Here, U(1)$''$ is generated by the linear combination of 
hypercharge and U(1)$'$ generators 
which leaves the Higgses invariant: 
\ben
Y'' = {Y'} - ({q_L + q_E})Y.
\een
There are still two Abelian symmetries, and SU(2) is completely broken with no discrete subgroup. Hence cosmic strings are not formed at this transition. 

We expect reheating to be very rapid 
\cite{GarciaBellido:1997wm,BasteroGil:1999fz,Felder:2000hj,GarciaBellido:2002aj,DiazGil:2005qp,Berges:2010zv}, 
as the period of oscillation of the fields is of order $M^{-1}$, which is much less than a Hubble time, and the couplings of the Higgs field are not all small.
Hence the universe 
regains a relativistic equation of state almost immediately,   
and thermalises at a temperature $T_\textrm{rh1}$ given by 
\ben
T_\textrm{rh1} = \left(\frac{30}{g_{\text{rh1}}\pi^2} \right)^\frac14 M = \left(\frac{30}{g_{\text{rh1}}\pi^2} \right)^\frac14 \sqrt{\la} s_c
\een
where $g_{\text{rh1}}$ is the effective number of relativistic degrees of freedom at temperature $T_\textrm{rh1}$.  
From (\ref{e:ScLim}), and taking $g_{\text{rh1}} = 915/4$ (a slight overestimate), we find 
\ben
T_\textrm{rh1} \simeq 2.2 \sqrt{\la} \times 10^{15} \; \text{GeV}.
\een
Hence the range of reheat temperatures corresponding to the range of couplings 
defined by \eqn{e:LamUppBou} and \eqn{e:LamLowBou}  is  
\ben
\label{e:Trh1Ran}
0.7  \times 10^{14} \lesssim T_\textrm{rh1}/\text{GeV}  \lesssim 5  \times 10^{14}.
\een
Finally, we note that large vevs of other fields along supersymmetric flat directions can lead to blocking of  particle production during reheating \cite{Allahverdi:2007zz}. On the other hand,  radiative corrections during inflation generically generate masses of order $y^2H^2$ \cite{Garbrecht:2006aw}, where $y$ is a combination of Yukawa couplings, and so we expect that other vevs besides that of the inflaton will be generally small.  We leave a detailed examination of the flat directions for another work, assuming for now that any flat directions which do not have $y$ of order 1 are small.

\subsection{High temperature ground state}

As the universe reheats, it will seek a minimum of the finite
temperature effective potential, or equivalently the free energy
density.  To discuss the free energy, it is convenient to define a dimensionful field 
$X = v_+\chi$, with $v_+ =  \sqrt{2}v_{\phi}= 2M/\sqrt{\la_1}$.  
The free energy density can then be expressed as 
\ben
f(X,T) = -\frac{\pi^2}{90}g_\textrm{eff}(X,T)T^4,
\een
where $g_\textrm{eff}(X,T)$ is the effective number of relativistic
degrees of freedom at temperature $T$.  At
weak coupling, $g_\textrm{eff}(X,T)$ can be calculated in the
high-temperature expansion for a particle of mass $m \ll T$
\cite{Dolan:1973qd}, 
\ben
g_\textrm{eff}(X,T) \simeq c_0 - c_1 \frac{90}{\pi^2} \frac{m^2}{T^2}, \quad 
\een
where there are contributions to $c_0$ of $1, \frac78$ and to 
$c_1$ of  $\frac{1}{24},\frac{1}{48}$ for bosons and fermions respectively. 
For particles with $m > T$, $g_\textrm{eff}$ is exponentially suppressed.

We can see that $X = 0$ is a local minimum for temperatures $m_{\frac{3}{2}} \ll T \lesssim M$, because away from that point 
the  U(1)$''$ gauge boson develops a mass proportional to $\vev{\phi}$, and so $g_\textrm{eff}$ decreases. 
For similar reasons $X_\phi = v_+\pi/2$ is also a local minimum: away from that point 
the \mssm\ particles develop masses and again reduce $g_\textrm{eff}$. 

In fact, by counting relativistic degrees of freedom at temperatures $m_{\frac{3}{2}} \ll T \lesssim M$ one finds that $X_\phi$ is the global minimum at high temperature. In the $h$-vacuum the relativistic species 
are the $\Phi, \Phib$ chiral multiplets and the U(1)$''$ gauge multiplet. In the $\phi$-vacuum, the particles of 
the \mssm\ are all light relative to $T$. Hence
\bea
f(0,T) &\simeq& - \frac{15}{2} \frac{\pi^2}{90}T^4, \\
f(X_\phi,T) &\simeq& -\frac{915}{4} \frac{\pi^2}{90}T^4.
 \eea
The minima of the free energy density are separated by a free energy barrier of height $\sim T^4$. 
The transition rate can be calculated in the standard way \cite{Linde:1978px} by calculating the free energy of
the critical bubble $E_c$. The transition rate per unit volume is then
\ben
\Ga \sim T^4 \left( \frac{E_c}{2\pi}\right)^\frac32 \exp\left(-\frac{E_c}{T}\right).
\een
The critical bubble is a solution to the equation 
\ben
X'' + \frac{2}{r}X' + V^T_\textrm{eff}(X) = 0,
\een
where $r$ is the radial distance from the bubble centre, and we have neglected O(1) complications in the 
kinetic term from the non-linear field transformation.
An order-of-magnitude estimate can be given, 
recognising that $X$ has to change by an amount $\De X \sim v_+$ from the inside to the outside of the 
bubble, while negotiating a local free energy bump of order $ \De V^T_\textrm{eff} \sim T^4$. 
Neglecting the damping term, one can translate the equation into a harmonic oscillator problem, finding that the critical bubble radius is approximately
\ben
r_c \sim \frac{\De X}{\sqrt{\De V^T_\textrm{eff}}},
\een
and so the critical bubble energy
\ben
E_c \sim \De V^T_\textrm{eff} r_c^3 \sim \frac{\De X^3}{\sqrt{\De V^T_\textrm{eff}}} \sim \frac{v_+^3}{T^2}.
\een
The universe will stay in the wrong ground state if the transition rate per unit volume is significantly below the Hubble rate per Hubble volume, or $\Ga < H^4$. 
Hence the reheat temperature $T_\textrm{rh1}$ should be parametrically
\ben
 T_\textrm{rh} < \frac{v_+}{[\ln(\mpl/v_+)]^\frac13}.
\een
Recalling that $T_\textrm{rh} \simeq M$ and $v_+ = 2M/\sqrt{\la_1}$, 
we see that if inflation exits to the $h$-vacuum it is likely that the universe continues to evolve with large (inflation-scale) Higgs vevs, provided $\la_1 \ll 1$.

\section{Review of gravitino constraints}

There are strong constraints on the gravitino mass and lifetime from
cosmology 
\cite{Weinberg:1982zq,Nanopoulos:1983up,Khlopov:1984pf,Ellis:1984eq}.
If the gravitinos are unstable, they can conflict with Big Bang
Nucleosynthesis (\bbn) by photodissociating light elements,  or they can
decay directly into the \lsp, which in turn produces a limit from the
known density of dark matter in the standard cosmological model.  The
gravitino may also be the \lsp, in which case the dark matter constraint
applies directly.

Gravitinos are produced by collisions of high-energy particles in the thermal bath, principally gluons and gluinos, with an abundance of approximately  \cite{Kawasaki:2008qe}
\ben
Y^\text{th}_{\frac32} \simeq \om_{\tilde G} \left(2.4 + 1.4 \frac{M^2_{\tilde g}}{\mgrav^2} \right)\times 10^{-13}   \left(\frac{T_\textrm{rh}}{10^{9}\; \textrm{GeV}}\right),
\label{e:GraAbu}
\een
where $M_{\tilde g}$ is the gaugino mass at the GUT scale. We include an O(1) factor $ \om_{\tilde G}$ to take into account the theoretical uncertainties \cite{Bolz:2000fu,Pradler:2006hh,Rychkov:2007uq}, arising from the strong dynamics of the coloured plasma.

\bbn\ constraints \cite{Kawasaki:2008qe} are not easily summarised,  but
are much tighter for lighter gravitinos which decay during or after 
nucleosynthesis, as relevant for the \cmssm.  For gravitino masses less
than about O(10) TeV, the reheat temperature is bounded above by
$T_\textrm{rh} \lesssim (0.2 - 1)\times 10^6$ GeV.  For higher gravitino
masses, the dark matter density provides a bound, and so it is
appropriate use \eqn{e:GraAbu} in the limit $M^2_{\tilde g}/\mgrav^2 \to
0$..

Given that the \lsp\ density parameter arising from a particular relic abundance in the \mssm\ 
is  
\ben
\Omega_{\rm \lsp}h^2 \simeq 2.8\times 10^{10}\frac{m_{\rm \lsp}}{100\,\textrm{GeV}}Y_{\frac32},
\een
the \lsp\ density parameter from (high mass) thermally produced gravitinos can be found as    
\ben \label{e:GraCon} \Omega_\textrm{\lsp}h^2
\simeq \om_{\tilde G}6 \times 10^{-3}  \frac{m_\textrm{LSP}}{100\; \textrm{GeV}} 
\left(\frac{T_\textrm{rh}}{10^{9}\; \textrm{GeV}} \right).
\een 
This must be less than or equal to the dark matter abundance inferred 
from the \cmb\ \cite{Komatsu:2010fb}
\ben
\Om_\textrm{dm}h^2 \simeq 0.11.
\een
The presence of cosmic strings in our model, although affecting the \cmb\ power spectrum, does not significantly affect this inferred value \cite{Urrestilla:2011gr}.

In our model, we will see that the gravitinos generated by the first
stage of reheating  are diluted by a period of thermal inflation. The
constraint therefore applies to reheating after thermal inflation.  We
will also see that the second reheat temperature is about $10^9$ GeV,
and so we can only tolerate unstable gravitinos of mass greater than
about $10$ TeV in order not to spoil \bbn.   This is natural in \amsb,
problematic in \gmsb, while the \cmssm\ keeps $\mgrav$ as a separate
parameter. 

There are also non-thermal production mechanisms from coherent
oscillations of the inflaton 
\cite{Nilles:2001ry,Nilles:2001fg} 
and from ordinary perturbative decay \cite{Kawasaki:2006gs}, whose rates
depend on the  inflaton mass and vev.   We will see in the next section
that the relevant inflaton mass and vev will be those of the Higgs. 
However the \bbn\ constraints mean that the gravitino, when it is not the \lsp, must be much more massive
than the Higgs and so cannot be produced by direct decays. Hence only
thermal production is relevant.

\section{Higgs thermal inflation and gravitinos}
\label{s:HigTheInfGra}

As the temperature falls, the energy density difference between the vacua becomes
comparable to thermal energy density, and the universe can seek its true ground state, which 
is $\chi = \pi/2$, the $\phi$-vacuum.

At zero temperature we can write the
difference in energy density between the $h$-vacuum and the
$\phi$-vacuum as (see Eqs. (\ref{eq:VdiffAMSB}), 
(\ref{eq:VdiffAMSBacc}), (\ref{eq:VdiffGMSB}) and (\ref{eq:VdiffmSUGRA}))
\ben
\De V_\textrm{eff}^0 \simeq v_+^2 m^2_\textrm{sb}
\een
where we recall that $v_+^2 = 4M^2/\la_1$, and we have defined
an effective \susy-breaking scale $m_\textrm{sb}$. In the supersymmetry-breaking scenarios under consideration  
\ben
\label{e:msbVal}
m^2_\textrm{sb} \simeq 
\left\{
\ba{cl}
m^2_\frac{3}{2}\left(\frac{g'^2}{16\pi^2}\right)^2 q^2_\phi {Q}, & (\textrm{AMSB}), \cr
\La_\text{s}^2 \frac{\la_1}{\la_3} \left[ \frac{3}{8} \left(  \frac{g^2_2}{16\pi^2} \right)^2 + \frac{1}{8} \left(  \frac{g^2_1}{16\pi^2} \right)^2  \right]& (\textrm{\gmsb}), \cr
  \half(m_0^2 -A^2/4)\left[\frac{\la_1}{\la_3} -1\right] & (\textrm{\cmssm}). \cr
\ea
\right.
\een
A period of thermal inflation \cite{Lyth:1995ka}
starts at 
\ben
T_\textrm{i} \simeq \left( \frac{30}{g_\textrm{i}\pi^2}  v_+^2m_\textrm{sb}^2\right)^\frac14 ,
\een
where $g_{\text{i}}$ is the effective number of degrees of freedom at temperature $T_\textrm{i}$.
The \cmb\ normalisation (\ref{e:ScLim}) for $N$ e-foldings of standard hybrid inflation gives $(v_+/\mpl) \simeq 5\times 10^{-3} (40/N)^\frac14$. 
Using the number of degrees of freedom for a $\textrm{U(1)}_\textrm{em}\otimes\textrm{U(1)}''$ theory with two light chiral multiplets $\Phi$ and $\Phib$, $g_\textrm{i} = 15$, we have (on dropping the unimportant dependence on $N$)
\ben
T_\textrm{i} \simeq 2.2 \times 10^{9} \left(\frac{m_\textrm{sb}}{1\;\textrm{TeV}}\right)^{\half} \; \textrm{GeV}.
\een
The $h$-vacuum must be a local maximum at zero temperature, i.e. the soft mass terms $m_\phi^2|\phi|^2 + m_{\phib}^2|\phib|^2$ must be negative. If the $h$-vacuum were a local minimum, one can estimate that the tunnelling rate per Hubble time per Hubble volume \cite{Coleman:1980aw} would be
\ben
\frac{\Ga}{H^4} \sim \frac{\msb^4\mpl^4}{M^8}e^{-S_E}, 
\een
where $S_E$ is the action of the Euclidean tunnelling solution. This ratio must be of order unity for the universe not to remain trapped in the false vacuum \cite{Guth:1981uk}, and since the prefactor is much less than unity, we see that we cannot allow a metastable $h$-vacuum for a graceful exit from thermal inflation.

Thermal inflation continues until 
the quadratic term in the thermal potential ${g'}^2T^2(|\phi|^2 + |\phib|^2)$ becomes the same size
as the negative soft mass terms. 
Near the false vacuum,  the high temperature effective potential for the field $X$ breaking the U(1)$''$  symmetry can be written \cite{Kirzhnits:1976ts,Linde:1978px} 
\ben
V_\textrm{eff}(X) \simeq \half \ga(T^2 - T_0^2)X^2 - \frac{1}{3}\de T X^3 + \frac{1}{4}\la_X X^4,
\een
where $\ga$, $\de$ and $\la_X$ are dimensionless constants, and $T_0 \simeq |m_\phi|/g'$. 
The cubic term arises from the gauge boson, and the 
transition is first order provided $\la_X < e^4$, where $e$ is the effective U(1)$''$ gauge coupling \cite{Kirzhnits:1976ts,Linde:1978px}.

Hence the transition which ends thermal inflation takes
place at $T_\textrm{e} \sim m_\textrm{sb}$, and the number of e-foldings
of thermal inflation is 
\ben
N_\th \simeq \half \ln \left( \frac{v_+}{m_\textrm{sb}}\right) \simeq 15 - \ln\left( \frac{m_\textrm{sb}}{1\;\textrm{TeV}}\right), 
\een
Thus gravitinos will be diluted to unobservably low densities, as will any baryon number generated prior to thermal inflation,  and any other dangerous GUT-scale relics such as monopoles.
 
After thermal inflation ends, there is another period of reheating as the energy of the modulus $X$ is converted to particles.  Around the true vacuum, the $X$ is mostly Higgs, and so its large amplitude oscillations will be quickly converted into the particles 
of the \mssm. The natural oscillation frequency around $\chi = \pi/2$ is of order $\mgrav$, while the Hubble rate is of order $\mgrav M/\mpl$. Hence in much less than an expansion time, the vacuum energy will be efficiently converted into thermal energy.  The reheat temperature following thermal inflation is thus 
\ben
T_\textrm{rh2} = \left( \frac{30}{g_\textrm{rh2}\pi^2} \De V_\textrm{eff}^0 \right)^\frac14 \simeq 0.5T_\textrm{i} \simeq 
1.1 \times 10^{9} \left(\frac{m_\textrm{sb}}{1\;\textrm{TeV}}\right)^{\half} \; \textrm{GeV},
\een
where $g_\textrm{rh2}$ is the effective number of relativistic degrees of freedom at $T_\textrm{rh2}$, given its \mssm\ value 
$g_\textrm{rh2}=915/4$.
This second reheating regenerates the gravitinos, and we may apply the gravitino density formula  \eqn{e:GraCon}, finding
\ben
\label{e:OmLSP}
\Omega_\textrm{\lsp}h^2
\simeq 6\times 10^{-3} \om_{\tilde G} \frac{m_\textrm{LSP}}{100\; \textrm{GeV}} 
\left(\frac{m_\textrm{sb}}{1\;\textrm{TeV}} \right)^{\half}
\een
We can convert the relic density into a constraint on the effective \susy-breaking scale $\msb$, requiring that the \lsp\ density is less than or equal to the observed dark matter abundance, $\Om_\textrm{dm}h^2 \simeq 0.11$. 
\ben
\msb \lesssim  3\times 10^{2}\frac{1}{\om^2_{\tilde G}}\left(\frac{m_\textrm{LSP}}{100\; \textrm{GeV}}\right)^{-2} \; \textrm{TeV}
\label{e:GravBouMsb}
\een
The parameter $\msb$ is directly related to physical observables differently in the different \susy-breaking schemes, for which we can derive constraints.

\subsection{Gravitino constraint in \amsb}

Using \eqn{e:GravBouMsb} 
and the expression for $m_{\rm sb}$ in \eqn{e:msbVal}, we find 
\ben
m_\frac32 \lesssim \frac{5\times 10^{4}}{{g'}^2q_{\Phi} \sqrt{Q}} \left(\om_{\tilde G} \frac{m_\textrm{LSP}}{100\; \textrm{GeV}}\right)^{-2}\;\textrm{TeV}.
\een
Hence \amsb-based models requires a high gravitino mass in order to saturate the bound and generate the dark matter.  

We can be a bit more precise if we use use the phenomenological 
relations derived in \cite{Hindmarsh:2012wh}.  
Firstly, in order to fit $\mu_h$ we have (using \eqns{eq:muans}{eq:hlaone})
\ben
q_{\Phi}^2 {g'}^2 \simeq  \frac{\la_1}{\la_3},
\een
while we can use a phenomenological formula for the \lsp\ mass 
\ben
m_\textrm{\lsp} \simeq 3.3\times 10^{-3} m_\frac32.
\een
Hence
\ben
m_\frac32 \lesssim {360} \left( \frac{1}{\om^2_{\tilde G}} \frac{q_{\Phi}}{\sqrt{Q}}\frac{\la_3}{\la_1} \right)^{\frac{1}{3}}\;\textrm{TeV}.
\een
We also have a constraint (\eqn{eq:la3range})
 on ${\la_3}/{\la_1}$ from requiring the exit to a false $h$-vacuum. 
Hence in order for the \lsp\ in this model to comprise all the dark matter, we have
\ben
  \left(\frac{1}{\om^2_{\tilde G}} \frac{q_{\Phi}}{\sqrt{Q}}\frac{q_H^2}{q_{\Phi}^2} \right)^{\frac{1}{3}} 
\lesssim \frac{m_\frac32}{360\;\textrm{TeV}} \lesssim 
\left(\frac{1}{\om^2_{\tilde G}} \frac{q_{\Phi}}{\sqrt{Q}} \right)^{\frac{1}{3}}.
\een
For example, taking $q_L=0$ as in \cite{Hindmarsh:2012wh}, we find that $m_\frac{3}{2}$ is independent of $q_E$ and in the range 
\ben
{130} \, \om_{\tilde G}^{-\frac23} \;\textrm{TeV}  \lesssim m_\frac32 \lesssim {250}\,  \om_{\tilde G}^{-\frac23}  \;\textrm{TeV},
\een
where we recall from the discussion around \eqn{e:GraAbu} that $\om_{\tilde G}$ is O(1). 
It was noted in \cite{Hindmarsh:2012wh} that a Higgs of mass 125 GeV
demands a gravitino mass of about 140 TeV in \samsb, which is compatible 
with an \lsp\ produced by gravitino decays being the dark matter.

\subsection{Gravitino constraint in \gmsb}

In the  \gmsb\ framework the \lsp\ is usually the gravitino, whose mass is given by 
\ben
\mgrav = \frac{1}{\sqrt{3}kN_{\rm mi}}\left( \frac{M_X}{\mpl}\right)  \La_\text{g},
\label{e:GravMasGMSB}
\een
where $k<1$ parametrises the fraction of the total F-term contained in the messenger sector, and we recall that $M_X$ is the messenger scale. 
It is more convenient to phrase the dark matter constraint 
in terms of the larger electroweak gaugino mass $M_2$, which dominates in the equation for the \susy-breaking scale $\La_\text{g}$, 
as in \eqn{e:MguaginoGMSB}, and therefore \eqn{e:msbVal} can be rewritten 
\ben
\msb^2 \simeq \frac{3}{8}\frac{\la_1}{\la_3} \frac{M^2_2}{{N_{\rm mi}}}.
\een
Hence
\ben
\left(\frac{\mgrav}{1\; \textrm{TeV}}\right)^2 \lesssim 
{5}{\sqrt{N_{\rm mi}}} \sqrt{\frac{\la_3}{\la_1}}\frac{1}{\om^2_{\tilde G}}\left(\frac{M_2}{1\; \textrm{TeV}}\right)^{-1}.
\een
The bound can be saturated for a TeV-scale gravitino with TeV-scale gaugino masses without special tuning of the ratio $\la_3/\la_1$. A lighter gravitino forces the gaugino mass upwards.

There is a separate constraint from decays of the \nlsp\ (which is generally a neutralino for unless $M_X$ is small), which may interfere with Big Bang Nucleosynthesis (see e.g. \cite{Giudice:1998bp}).
A careful analysis of the nucleosynthesis constraints \cite{Gherghetta:1998tq} shows that a messenger mass of up to about $10^{14}$ GeV is allowed, before hadronic jets injected after $10^4$ s  results in the overproduction of $^7$Li.  
The combination of the dark matter and \bbn\ constraints $M_X \lesssim 10^{14}$ GeV may be written
\ben
\left(\frac{\mgrav}{1\; \textrm{TeV}}\right)^3 \lesssim 
{0.5}{\sqrt{N_{\rm mi}}} \sqrt{\frac{\la_3}{\la_1}}\frac{1}{\om^2_{\tilde G}}.
\een

\subsection{Gravitino constraint in the \cmssm}

In the \cmssm, the gravitino bound \eqn{e:GravBouMsb} 
can be expressed in terms of the soft scalar masses $m_0$ and the trilinear parameter $A$ from 
\eqn{e:msbVal}, as
\ben
\left(m_0^2 - {A^2}/{4}\right)^{\half} \lesssim 
{5\times 10^{2}} \left(\frac{\la_1}{\la_3} -1 \right)^{-\half} \left(\om_{\tilde G} \frac{m_\textrm{LSP}}{100\; \textrm{GeV}}\right)^{-2}\;\textrm{TeV}.
\een
This is a very weak bound, unless the ratio $\la_3/\la_1$ is very small:
hence there is generically a very low density of \lsp\ dark matter
generated by decays of gravitinos. Instead,  the \cmssm\ can generate an
acceptable dark matter density through the standard freeze-out scenario
\cite{Ellis:1983ew} (see
\cite{Bechtle:2012zk,Fowlie:2012im,Buchmueller:2012hv} 
for recent analyses of the the \cmssm\ parameter space in the light of
recent Higgs results). This requires that the gravitino mass is larger
than about 10 TeV to avoid \bbn\ constraints \cite{Kawasaki:2008qe}. We
conclude that the Higgs thermal inflation solution generally has no
effect on the gravitino problem in the \cmssm, beyond determining the
reheat temperature and hence the standard \bbn-induced lower bound
gravitino mass.

\section{Conclusions}

In this paper we have shown how models which couple F-term hybrid
inflation with the \mssm\ without extra ingredients naturally realise a
period of thermal inflation, with a reheat temperature of around $10^9$
GeV, while generating the Higgs $\mu$-term. The inflation is
driven by the relaxation of the Higgs fields to zero in a potential
generated by the Higgs and waterfall field soft terms. This second
period of intermediate scale inflation, which we have called Higgs
thermal inflation, has a number of beneficial effects.  It solves the gravitino
overabundance problem of supersymmetric cosmology, while still
maintaining the possibility of leptogenesis.  It reduces the cosmic
string mass per unit length so that \cmb\ bounds are satisfied, and
renders it independent of the inflaton couplings. Hence the scalar
spectral index is not driven to unity in the effort to make the strings
light, from which the tight constraints on standard F-term hybrid
inflation are generated \cite{Battye:2010hg}.  The period of thermal
inflation means a reduced number of e-foldings of F-term inflation are
required, and the scalar spectral index is reduced:  in the range of
inflaton couplings where the inflaton potential is dominated  by the
radiative corrections, we find $n_s \simeq 0.976(1)$.

The \mhissm\ is  the simplest, attractive,  formulation of this
scenario, and generates right-handed neutrino masses as well as the
Higgs $\mu$-term.  We found constraints on the couplings and soft terms
in order for the scenario to work: i.e.\ for F-term inflation to exit
towards a vacuum with inflation-scale vevs for the Higgs field, and for
that vacuum to be unstable. We investigated the implications of these
constraints in three popular supersymmetry-breaking scenarios: \amsb\
(where the model coincides with strictly anomaly-mediated
supersymmetry-breaking \cite{Hindmarsh:2012wh}), \gmsb,  and the \cmssm.
 We found constraints on the ratio of the inflaton  couplings in \amsb\
(\eqn{e:EWCon}) and \gmsb\ (\eqn{e:gmsb}),  and that in the \cmssm\ the
soft scalar mass  must be greater than the half the magnitude of the
soft trilinear term, multiplied by the square root of the ratio of the
inflaton  couplings (\eqn{eq:sugrahigg}).

In \amsb, the gravitino problem becomes the gravitino solution:  the
observed dark matter density can be generated by the decays of
gravitinos which are produced thermally following Higgs thermal
inflation.  In \gmsb, the gravitino is the \lsp, and a weak upper bound
on its mass of about 1 TeV follows from the combined requirement that it
supply the dark matter without \nlsp\ decays spoiling nucleosynthesis. 
In the \cmssm, the density of thermally-produced gravitinos is generally
sub-dominant, and the standard freeze-out scenario must do the work of
making neutralino dark matter.  However, the gravitinos must decay early
enough not to spoil nucleosynthesis, meaning that the gravitino mass must
be O(10) TeV or greater.

A reheat temperature of $10^9$ GeV is broadly consistent with thermal
leptogenesis, provided at least one right-handed neutrino is light
enough to be thermally produced.  We leave the details for a future
publication. 

The \mhissm\ predicts the formation cosmic strings, with dimensionless
mass per unit length estimated as $G\mu_\text{s} \simeq 10^{-7}$
\cite{Hindmarsh:2012wh}. While satisfying current \cmb\ bounds,  there are
tight bounds on the GeV-scale cosmic $\ga$-ray spectrum
\cite{Bhattacharjee:1997in}, so strings should have a very small
branching fraction into $\ga$. Strings may instead decay into
gravitational waves, but there are also increasingly strict bounds on
the stochastic gravitational wave background from pulsar timing 
\cite{Battye:2010xz,Kuroyanagi:2012wm}.   
Should the bounds ultimately fall below the predicted value of 
$G\mu_\text{s}$, this will rule out the \mhissm\, but not the Higgs
thermal inflation scenario in general, which remains a possibility
whenever the inflaton is coupled to a set of waterfall fields which
include the Higgs.

\acknowledgments

This research was supported in part by the Science and Technology
Facilities Council [grant numbers ST/J000477/1 and ST/J000493/1],   the
Project of Knowledge Innovation Program (PKIP) of Chinese Academy of
Sciences, Grant No. KJCX2.YW.W10, and the National Science Foundation 
under Grant No. NSF PHY11-25915. MH and TJ  gratefully acknowledge the
hospitality of the  Kavli Institutes for Theoretical Physics, in  China
and Santa Barbara respectively, and also the role of the UK Particle Cosmology
workshop in the development and dissemination of this research.

\appendix

\section{\samsb\ soft parameters and inflaton coupling constraints}

In this appendix we give formulae for a more accurate calculation of the \samsb\ soft parameters  in 
Section \ref{s:TruMinAMSB}, 
relevant for the calculation of the vacuum energies and hence the  constraint on $\la_3/\la_1$.

The $h_{\lambda_1}$ and $h_{\lambda_3}$ trilinear soft terms are determined in accordance with 
\eqn{eq:AD2}:
\bea
h_{\lambda_1} &=& - m_{\frac{3}{2}}\frac{\la_1}{16\pi^2}\left(
3\la_1^2 +\frac{1}{2}\Tr \lambda_2^2 +2\lambda_3^2 - 4q_{\Phi}^2{g'}^2 \right), \\
h_{\lambda_3} &=& - m_{\frac{3}{2}}\frac{\la_3}{16\pi^2}(
\Tr Y_E Y_E^{\dagger} + 3\Tr Y_D Y_D^{\dagger}+ 3\Tr Y_U Y_U^{\dagger} 
+\lambda_1^2 + 4\Tr \lambda_3^2\nn
&& -  3g_2^2 - g_1^2 - 4q_{H}^2{g'}^2), 
\eea
while the mass soft terms are
\bea
m_\phi^2 & = &  \mgrav^2 \frac{1}{32\pi^2} \mu \frac{d}{d\mu}\left( \la_1^2 + \half \Tr\la_2^2  - 2{g'}^2q_{\Phi}^2\right),\\
m_{\phib}^2 & = & \mgrav^2 \frac{1}{32\pi^2} \mu \frac{d}{d\mu}\left( \la_1^2   - 2{g'}^2q_{\Phi}^2\right), \\
m_{h_1}^2 & = &  \mgrav^2 \frac{1}{32\pi^2} \mu \frac{d}{d\mu}
\left( \la_3^2 + \Tr Y_E Y_E^{\dagger} + 3\Tr Y_D Y_D^{\dagger}  - 2{g'}^2q_{H}^2 - \half g_1^2 - \frac{3}{2} g_2^2\right), \\
m_{h_2}^2 & = & \mgrav^2 \frac{1}{32\pi^2} \mu \frac{d}{d\mu}
\left( \la_3^2 + \Tr Y_N Y_N^{\dagger} + 3\Tr Y_U Y_U^{\dagger}  - 2{g'}^2q_H^2 - \half g_1^2 - \frac{3}{2} g_2^2\right).
\eea
In the following, we include the \sm\ gauge couplings and the top Yukawa coupling, which we shall denote 
$y_t$.  We also retain the neutrino Yukawas $Y_N$, since 
their magnitude is model dependent. 
We will assume that the value of $\tan\beta$ is such that  
it is a good approximation to neglect all the other Yukawas,
and we will also neglect $\la_1$ and $\la_3$.  

The relevant soft
breaking parameters are then 
\bea
h_{\lambda_1} &=&  m_{\frac{3}{2}}\frac{\la_1}{16\pi^2}\left(
4q_{\Phi}^2{g'}^2 - \frac{1}{2}\Tr \lambda_2^2   \right), \\
h_{\lambda_3} &=& m_{\frac{3}{2}}\frac{\la_3}{16\pi^2}\left(
4q_{H}^2{g'}^2 +  3g_2^2 + g_1^2 - 3y_t^2  -\Tr Y_N^2  \right), \\
m_\phi^2 & = &  - \frac{\mgrav^2}{(16\pi^2)^2}\left[ 2q_{\Phi}^2Q{g'}^4 - \Tr \la_2^4 \right . \nn
&& \qquad\qquad - \left. \half \Tr\la_2^2\left\{ \half\Tr\la_2^2 + 4\Tr Y_NY_N^\dagger -2(q_{\Phi}^2 + 2 q_N^2){g'}^2 \right\}  \right],\\
m_{\phib}^2 & = & - \frac{\mgrav^2}{(16\pi^2)^2} \left[2q_{\Phi}^2Q {g'}^4\right],\\
m_{h_1}^2 & = & - \frac{\mgrav^2}{(16\pi^2)^2} \left[2q_{H}^2Q{g'}^4  
+ \frac{11}{2}g_1^4  + \frac{3}{2}g_2^4 \right],\\
m_{h_2}^2 & = & - \frac{\mgrav^2}{(16\pi^2)^2} 
\left[2q_{H}^2Q{g'}^4 + \frac{11}{2}g_1^4 + \frac{3}{2}g_2^4 
-6\Tr Y_N^4 - 2\Tr Y_N^2 \la_2^2\right. \nn
&& -2\Tr Y_N^2\left\{\Tr Y_N^2 +3y_t^2- 3g_2^2 
- g_1^2 - 2(q_N^2 + q_L^2 + q_H^2){g'}^2\right\}\nn
&& \left. - 3y_t^2\left(
6y_t^2 + \Tr Y_N^2 - \frac{16}{3}g_3^2 - 3g_2^2 - \frac{13}{9}g_1^2
-2(q_Q^2 + q_{t^c}^2 + q_H^2){g'}^2\right)
\right].
\eea
We have assumed above for simplicity that, like $\la_2$, 
$Y_N$ is real and diagonal.

The difference in the energies between the large-Higgs and large-$\phi$ vacua can be written
\bea
\frac{(16\pi^2)^2 \la_1}{\mgrav^2M^2}(V_{h} - V_\phi) &\simeq& 
4Q q_{\Phi}^2{g'}^4 + \De_{m^2_\phi} + \half\left( 4q_{\Phi}^2{g'}^2 + \De_{h_1} \right)^2 \nn
&&   - \frac{\la_1}{\la_3}\left[4Q q_{H}^2{g'}^4 + \De_{m^2_h} + \half\left(4q_{H}^2{g'}^2 + \De_{h_3}\right)^2\right],
\label{eq:VdiffAMSBacc}
\eea

where
\bea
\De_{m_\phi^2} &=&  - \half \Tr\la_2^2\left( \half \Tr \la_2^2 
+ 4\Tr Y_NY_N^\dagger - 2(q_{\Phi}^2 + 2q_N^2){g'}^2\right)  -\Tr \la_2^4  \\
\De_{h_1} &=&  -\half\Tr \la_2^2\\
\De_{m_h^2} &=& 11 g_1^4  +  3 g_2^4 
-6\Tr Y_N^4 - 2\Tr Y_N^2 \la_2^2\nn
&-2&\Tr Y_N^2\left\{\Tr Y_N^2 +3y_t^2- 3g_2^2 
- g_1^2 - 2(q_N^2 + q_L^2 + q_H^2){g'}^2\right\}\nn
&-& 3y_t^2\left(6y_t^2 - \frac{11}{3}g_3^2 - 3g_2^2 - \frac{13}{9}g_1^2
-2(q_Q^2 + q_{t^c}^2 + q_H^2){g'}^2\right) \\
\De_{h_3} &=& g_1^2 + 3g_2^2 - 3|y_t|^2-\Tr Y_N^2
\eea
The condition on the couplings deriving from the vacuum energies can therefore be written
\ben
\frac{\la_3}{\la_1} >  \frac{q_H^2}{q_\phi^2} \frac{1 + (q_H^2/Q)\left(\be^2_H\De_{m^2_h} + 2\left(1+\half\be_H\De_{h_3}\right)^2\right)}
{1 + (q_\phi^2/Q)\left(\be_\phi^2\De_{m^2_\phi} + 2\left( 1 + \half\be_\phi\De_{h_1} \right)^2\right)},
\een
where
\ben
\be_H = \frac{1}{2q_H^2{g'}^2}, \quad \be_\phi = \frac{1}{2q_\phi^2{g'}^2}.
\een
Taking the values of the \sm\ couplings at the U(1)$'$ breaking scale to be the values at gauge coupling unification, we find using the renormalisation group analysis of 
\reference{Hindmarsh:2012wh}
\ben
g_1 \simeq 0.55, \quad g_2 \simeq 0.71, \quad g_3 \simeq 0.70; \quad y_t \simeq 0.51
\een
(where we have taken $\tan\beta = 16$). 
Hence, at this level of accuracy, 
\bea
\De_{m_h^2} &\simeq&  3.5 - 6\Tr Y_N^4 - 2\Tr Y_N^2 \la_2^2\nn
&-2&\Tr Y_N^2\left\{\Tr Y_N^2 -1.0 - 2(q_N^2 + q_L^2 + q_H^2){g'}^2\right\}\nn
&+& 1.6\left(q_Q^2 + q_{t^c}^2 + q_H^2\right){g'}^2 \\
\De_{h_3} &\simeq& 1.0 - \Tr Y_N^2 .
\eea
We can derive successive approximations. Firstly, neglecting terms of order $q_H^2/Q$ and $q_\phi^2/Q$, which is a good approximation given 
\eqn{e:Qeqn}, we have
\ben
\frac{\la_3}{\la_1} >  \frac{q_H^2}{q_\phi^2}.
\een
Secondly, we can neglect terms of order $\be_{H,\phi}$ and higher (which is not necessarily a good approximation), to obtain
\ben
\frac{\la_3}{\la_1} >  \frac{q_H^2}{q_\phi^2}\frac{1 + 2(q_H^2/Q)}
{1 + 2(q_\phi^2/Q)}.
\een
In the case $q_L = 0$, we obtain \eqn{e:EWConb}. 

We can also expand in powers of $\be_{H,\phi}$ while still neglecting terms of order $Y_N^2$, and bearing in mind that an acceptable electroweak vacuum requires $\Tr\la_2^2 \simeq 4\be_H$ \cite{Hindmarsh:2012wh}), we find to second order 
\ben
\frac{\la_3}{\la_1} >  \frac{q_H^2}{q_\phi^2}\frac{1 + 2(q_H^2/Q)\left(1 + [1.4+ 0.4(q_\phi^2+q_N^2)/q_H^2]\be_H + 2.0\be_H^2 \right)}
{1 + 2(q_\phi^2/Q)\left(1 + \frac14 \Tr\la_2^2[1+2q_N^2/q_\phi^2]- \half \be_\phi\Tr\la_2^2\right)}.
\een

\bibliographystyle{jhep}
\bibliography{fterm}

\end{document}